\documentclass[a4paper,final]{appolb}

\newcount\eLiNe\eLiNe=\inputlineno\advance\eLiNe by -1

\usepackage{hyperref}
\usepackage{pstricks,pst-node,pst-text,pst-3d}
\usepackage{colordvi}
\usepackage{epsfig}
\usepackage{rotating}
\usepackage{cite}

\newcommand{\be}{\begin{equation}}
\newcommand{\ee}{\end{equation}}
\newcommand{\bea}{\begin{eqnarray}}
\newcommand{\eea}{\end{eqnarray}}
\newcommand{\bqa}{\begin{eqnarray}}
\newcommand{\eqa}{\end{eqnarray}}

\begin{document}

\title{
{ \begin{flushleft}
DESY 07--183\\
SFB/CPP--07--70 \\ 
\vspace*{0.5cm}
\end{flushleft}}
FERMIONIC NNLO CONTRIBUTIONS TO BHABHA SCATTERING
\thanks{Presented by T.R.  at XXXI Conference of Theoretical Physics
``Matter to the Deepest: Recent Developments In Physics
of Funda\-men\-tal Inter\-\-actions'', Ustro\'n, Poland, 5-11 September 2007 \cite{transp}.
\\
Work supported in part by
Sonderforschungsbereich/Transregio TRR 9 of DFG
``Computergest\"utzte Theoretische Teilchenphysik",  by
the Sofja Kovalevskaja Award of the Alexander von Humboldt Foundation
  sponsored by the German Federal Ministry of Education and Research,
and by
the European Community's Marie-Curie Research Training Networks
MRTN-CT-2006-035505 ``HEPTOOLS'' and MRTN-CT-2006-035482 ``FLAVIAnet''.
}}

\author{S. Actis$^{a}$, M. Czakon$^{bc}$, J. Gluza$^{c}$, T. Riemann$^{a}$
\address{$^{a}$ Deutsches Elektronen-Synchrotron DESY\\
Platanenallee 6, D--15738 Zeuthen, Germany}
\address{$^{b}$ Institut f\"ur Theoretische Physik
und Astrophysik, Universit\"at W\"urzburg\\
Am Hubland, D-97074 W\"urzburg, Germany}
\address{$^{c}$ Institute of Physics, Univ. of
    Silesia, Universytecka 4, 40007 Katowice, Poland}
}
\maketitle
\begin{abstract}
We derive the two-loop corrections to Bhabha scattering from heavy fermions using dispersion relations. 
The double-box contributions are expressed by three kernel functions.
Convoluting the perturbative kernels with fermionic threshold functions or with hadronic data  allows to determine numerical results for small electron mass $m_e$, combined with arbitrary values of the fermion mass $m_f$ in the loop, $m_e^2<<s,t,m_f^2$, or with hadronic insertions.
We present numerical results for $m_f = m_{\mu}, m_{\tau}, m_{top}$ at typical small- and large-angle kinematics ranging 
from 1 GeV to 500 GeV. 
\end{abstract}

\section{Introduction}
Bhabha scattering is one of the theoretically best studied scattering processes at $e^+e^-$ colliders and can also be measured with a high precision.
The accuracy of the Monte-Carlo programs developed originally for physics at LEP is about $10^{-3}$, and with a complete two-loop calculation one may reach $10^{-4}$.
The latter number is indicative of efforts for the International Linear Collider (ILC), here especially in the GigaZ option running at the $Z$ boson resonance, but also for meson factories running at much smaller energies of about 1 or 10 GeV.

Recent years brought considerable progress in the determination of the virtual NNLO corrections. 
The virtual $O(\alpha^2)$ contributions to the massless differential Bhabha cross section have been determined in \cite{Bern:2000ie}.
Shortly after, this result was used for deriving the $O(\alpha^2 L)$  ($L=\ln(s/m_e^2)$) corrections to massive Bhabha scattering in \cite{Glover:2001ev}.
The missing  photonic correction terms of order $O(\alpha^2 L^0)$ were derived, also from \cite{Bern:2000ie}, in \cite{Penin:2005kf,Penin:2005eh}.
The virtual corrections from fermion loop insertions, including the corresponding  double-box diagrams, could not be covered by that method.
For $n_f=1$, \ie the case of only electron loops, the corresponding diagrams were evaluated analytically  in \cite{Bonciani:2003te,Bonciani:2003ai,Bonciani:2003cj,Czakon:2004wm},
and the net $n_f=1$ cross section in \cite{Bonciani:2004gi,Bonciani:2004qt}.

At this stage the numerically most important two-loop corrections were known.
For a complete treatment one needs additionally the $n_f=2$  two-loop corrections with heavy fermion insertions, including the hadronic corrections which replace the loop insertions from light quarks.
The leptonic $n_f=2$ contributions have been derived quite recently in two papers in the limit 
$m_e^2<<m_f^2<<s,t$; with a direct Feynman diagram calculation in \cite{Actis:2007gi}, and using  
a factorization formula that relates massless and massive amplitudes in \cite{Becher:2007cu} (for that method see also \cite{Mitov:2006xs}).

It might be interesting to mention that the original expectations on the necessity of a complete, direct two-loop massive Feynman diagram evaluation were not fulfilled.
After the analytical evaluation of a massive planar and a massive non-planar double-box diagram (both with seven propagators) in \cite{Smirnov:1999gc} and in \cite{Tausk:1999vh}, resp., there was hope to evaluate all the remaining diagrams soon.
There are 33 two-loop box master integrals, nine of them with seven lines \cite{Czakon:2004tg}.
In fact, from recent studies on the exact and mass expanded treatment of two-loop box master integrals in \cite{Heinrich:2004iq} and \cite{Czakon:2006pa}, resp., it became clear that an evaluation of all the massive master integrals is a more complicated task than was expected.
Quite recently, the case of non-planar master integrals was successfully treated in another, but related context \cite{Czakon:2007ej}. 
Proceeding similarly for Bhabha scattering seems feasible now.
As a matter of fact, due to these reasons, the direct Feynman diagram approach was not used for the phenomenologically needed two-loop predictions and the above-mentioned papers \cite{Smirnov:1999gc}-\cite{Czakon:2006pa} remained so far a mere interesting, challenging theoretical development. 

In this paper, we report on the evaluation of the leptonic $n_f=2$ two-loop contributions with {\em arbitrary mass} of the heavier fermion, \ie exploring the extended kinematical region $m_e^2<<m_f^2,s,t$.
We use the dispersion approach, so that our formulae may be applied without further modification also to hadronic corrections.

For a review on the status of Monte-Carlo studies for Bhabha scattering at this conference we refer to \cite{Montagna:us2007}, and for a discussion on progress related to radiative loop corrections to \cite{Kajda:us2007}.

\section{Formulae}
\begin{figure}
\begin{center}
\includegraphics[scale=0.5]{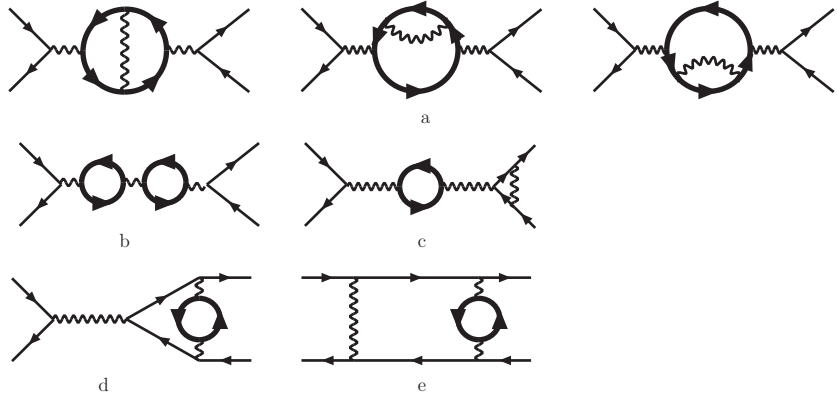}
\end{center}
\caption
{\label{fig-nf2}
Classes of  {two-loop diagrams} for Bhabha scattering containing at least one fermion loop.
}
\end{figure}

\begin{figure}
\begin{center}
\includegraphics[width=0.5\textwidth]{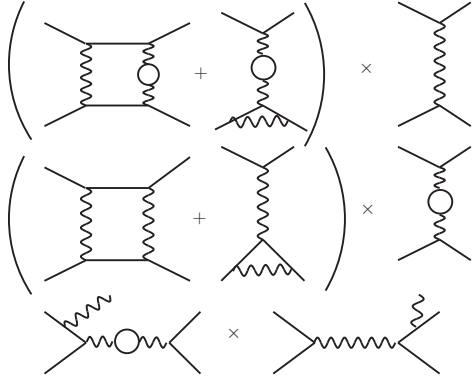}
\end{center}
\caption
{\label{fig-IRfin}
The fermionic two-loop boxes combine with other diagrams to an infrared-finite cross-section contribution.
}
\end{figure}

The classes of two-loop $n_f=2$ corrections  are shown in Figure \ref{fig-nf2}.
The four direct and four crossed fermionic two-loop box diagrams, obtained applying
proper permutations to the sample diagram shown in Figure \ref{fig-nf2}, are infrared (IR) 
divergent, and they have to be combined with other IR-divergent factorizable
corrections in order to get an IR-finite contribution to the cross section.
In particular, we have to add the interference of two-loop box (class $\rm e$) and reducible vertex
diagrams (class $\rm c$) with the tree-level amplitude to the interference of 
one-loop vertex and box diagrams with one-loop vacuum polarization  diagrams.
Finally, we construct an IR-finite quantity taking into account also the real emission of one soft photon
from one-loop vacuum polarization diagrams.
Sample contributions are given in Figure \ref{fig-IRfin}.

The net contribution of pure self-energy corrections (classes $\rm a-b$), irreducible vertex
diagrams (class $\rm d$), and the aforementioned IR-divergent contributions reads as
\begin{equation}
\frac{d \sigma^{\rm NNLO, ferm.}}{d \Omega}=
\frac{d \sigma^{\rm a,b}}{d \Omega} +
\frac{d \sigma^{\rm d}}{d \Omega} +
\frac{d \sigma^{\rm rest}}{d \Omega},
\end{equation}
where $d \sigma^{\rm rest}\slash d \Omega$ can be splitted in two components,
\begin{equation}
\frac{d \sigma^{\rm rest}}{d \Omega} =
\frac{d \sigma^{\rm box}}{d \Omega} +
\frac{d \sigma^{\rm fact.}}{d \Omega}.
\end{equation}

We concentrate now on the renormalized two-loop box diagrams of class \rm{e}, whose total
contribution to the cross section may be written as
\begin{eqnarray}
\frac{d \sigma^{\rm{box}}}{d \Omega}\,=\,
\left( \frac{\alpha}{\pi} \right)^2\,
\frac{\alpha^2}{2\, s}\,
\Bigl(\,
\frac{m_e^2}{s}\, {\rm Re}\, A_s \, +\,
\frac{m_e^2}{t}\, {\rm Re}\, A_t
\,\Bigr).
\end{eqnarray}
Here the auxiliary functions $A_s$ and $A_t$ can be conveniently expressed through three
independent form factors $B_{\rm{I}}$, with $\rm{i}=\rm{A},\rm{B},\rm{C}$,
evaluated with different kinematical arguments,
\begin{eqnarray}
\label{split}
A_s  &=& {B_{\rm A}(s,t)}\, +\, B_{\rm B}(t,s)\, + \, B_{\rm C}(u,t)\,-\, B_{\rm B}(u,s),
\nonumber \\
A_t  &=& {B_{\rm B}(s,t)}\, + \, B_{\rm A}(t,s)\, - \,B_{\rm B}(u,t)\,+\, B_{\rm C}(u,s).
\end{eqnarray}
The particular contribution of the diagram of Figure \ref{fig-nf2}
coming from the interference with the tree-level s-channel is $B_{\rm A}(s,t)$,
and from the t-channel is $B_{\rm B}(s,t)$.

For the evaluation of hadronic corrections, we observe that
each term of Eq.~(\ref{split}) can be written through the convolution of a kernel
function $K_{\rm I}$, $I=A,B,C$, with the hadronic cross-section ratio $R_{had}$, \begin{equation}
B_{{\rm I},had}(s,t) = 
\int_{4M_{\pi}^2}^{\infty}\, \frac{dz}{z}\, R_{had}(z)\, K_I(s,t,z) .
\end{equation}
For leptons and the top quark, we have to replace $4\,M_{\pi}^2 \to 4\,m_f^2$ and $R_{had} \to R_{fer}$, given by
\begin{eqnarray}
R_{fer}(z) = Q_f^2\, C_f\,\sqrt{1-4\, \frac{m_f^2} {z}}\left(1+2\, \frac{m_f^2}{z}\right) + \epsilon ~ R_{fer}^{\epsilon}(z),
\end{eqnarray}
where $Q_f$ is the electric charge in units of $|e|$ and $C_f$ is the color factor.
Being the box diagrams IR divergent, and showing poles in $\epsilon$, one should take into
account higher orders in $\epsilon$ for $R$. However, after assembling box diagrams
with factorizable corrections, IR poles cancel and $R$ can be evaluated at order $\epsilon^0$.

\begin{figure}
\begin{center}
\includegraphics[width=0.99\textwidth]{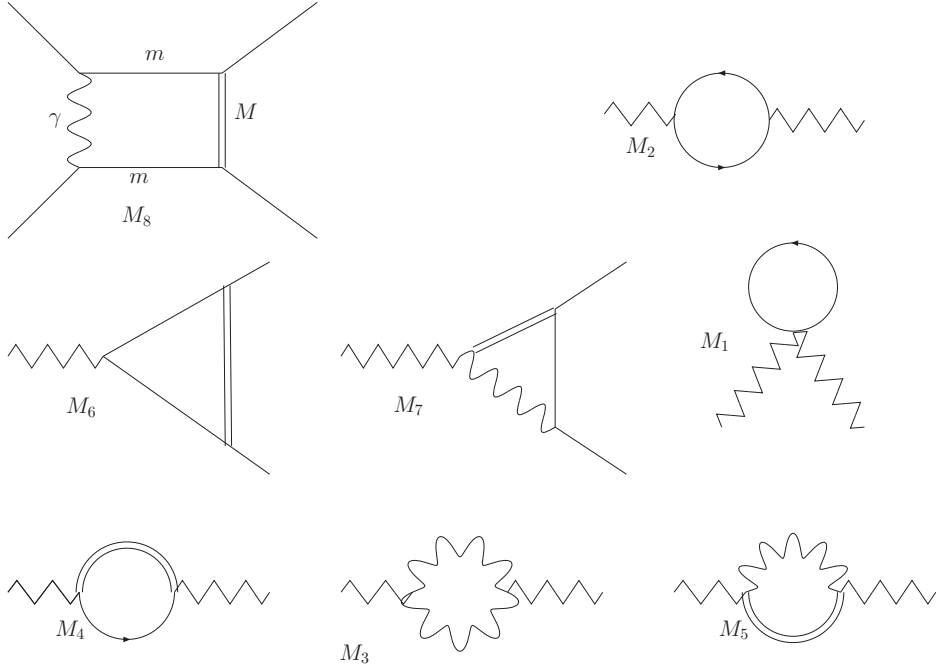}
\end{center}
\caption
{\label{fig-1loopmasters}
The master integrals for the two-loop box kernel functions. 
}
\end{figure}

The three box kernels are our main technical result.
They have been derived  with the aid of the master integrals of Figure \ref{fig-1loopmasters}, 
in the limit $m_e^2<<m_f^2, s,t$.
The master integrals were determined with {\ttfamily IdSolver} and evaluated with the Mathematica packages {\ttfamily ambre} \cite{Gluza:2007rt}  and {\ttfamily MB} \cite{Czakon:2005rk}, and eventually mass expanded with a Mathematica package. We also made use of {\ttfamily FORM} \cite{Vermaseren:2000nd}.
We reproduce one of the kernels here,\footnote{Some formulae have to be omitted here due to limited space; they may be found at the webpage  http://www-zeuthen.desy.de/theory/research/bhabha/bhabha.html.}
\begin{eqnarray}
K_C(x,y,z) &=&
\frac{1}{3\,m_e^2\,(y-z)} \, \Bigl\{
     2\,\frac{F_\epsilon}{\epsilon}\,x^2\,L_x
+ 4\,\zeta_2\,x^2\,\Bigl(\frac{z}{y}-2\Bigr)
- 2\,(x^2+y^2 \nonumber\\ &+&x\,y )\,L_x
 + x^2\,\Bigl(\frac{z}{y}-1\Bigr)\,L_y
 + 2\,x^2\,\Bigl(\frac{z}{y}-1\Bigr)\,L_y^2
 + 4\,x^2\,L_x\,L_y \nonumber\\
&+& x^2\,\Bigl(\frac{z}{y}-1\Bigr)\,\ln\Bigl(\frac{z}{m_e^2}\Bigr)
 - 2\,x^2\,\Bigl(\frac{z}{y}-\frac{1}{2}\Bigr)\,
     \ln^2\Bigl(\frac{z}{m_e^2}\Bigr)
+ 4\,x^2\,\Bigl(\frac{z}{y} \nonumber\\
&-&1\Bigr)
\, \ln\Bigl(\frac{z}{m_e^2}\Bigr)\,
     \ln\Bigl(1-\frac{z}{y}\Bigr)
 + 2\,x^2\,\ln\Bigl(\frac{z}{m_e^2}\Bigr)\,L_x
 - x^2\,\Bigl(\frac{z}{y}+\frac{y}{z}\nonumber\\
&-&2\Bigr)
\,   \ln\Bigl(1-\frac{z}{y}\Bigr)
- 4\,x^2\,\ln\Bigl(1-\frac{z}{y}\Bigr)\,L_x
 +  4 \,x^2\,\Bigl(\frac{z}{y}-1\Bigr)\,
     \textrm{Li}_2\Bigl(\frac{z}{y}\Bigr)\nonumber\\
&-& 2\,x^2\,\textrm{Li}_2\Bigl(1+\frac{z}{x}\Bigr)
\Bigr\},
\end{eqnarray}
where $L_x=\ln (-m_e^2\slash x)$, $L_y=\ln (-m_e^2\slash y)$ and $F_\epsilon$
is the normalization factor 
\begin{equation}
 F_\epsilon\, = \,
\, \left(\, \frac{m_e^2\, \pi\, e^{\gamma_E}}{\mu^2 }\, \right)^{-\epsilon}.
\end{equation}
Here $\mu$ is the 't Hooft mass unit and $\gamma_E$ is the Euler-Mascheroni
constant.

\section{Numerical Results}
The sum of the box contributions 
 with  IR-divergent
factorizable corrections (see Figure \ref{fig-IRfin} for sample cases)
is infrared-finite and can be cast in the  following form,
\begin{eqnarray}
\label{TotRemainder}
\frac{ d\sigma^{\rm rest} }{ d\Omega }&=& \left(\frac{\alpha}{\pi}\right)^2\, \frac{\alpha^2}{s}\,\Bigl\{
\int_{4M^2}^{\infty}\, dz\, \frac{R(z)}{z}\, \frac{1}{t-z}\, F_1(z)\nonumber\\
&+&
\textrm{Re}\,\int_{4M^2}^{\infty}\, dz\, \frac{R(z)}{z}\, \frac{1}{s-z+i\,\delta}\,
\Bigl[ \,F_2(z)\,+\, F_3(z)\, \ln\Bigl(1-\frac{z}{s+i\, \delta}\Bigr)\, \Bigr]\nonumber\\
&+&\pi\, \textrm{Im}\, \int_{4M^2}^{\infty}\, dz\, \frac{R(z)}{z}\, 
\frac{1}{s-z+i\,\delta} \, F_4(z) \Bigr\}.
\end{eqnarray}
We may show here one of the  auxiliary functions $F_i(z)$,
\begin{eqnarray}
F_1(z)&=& \frac{1}{3}\,\Bigl\{\,
\Bigl[\, 3\,\Bigl( \frac{t^2}{s}+2\,\frac{s^2}{t} \Bigr)+9\,\Bigl(s+t\Bigr)\Bigr]\, 
\ln\Bigl(\frac{s}{m_e^2}\Bigr)
+\Bigl[-z^2\Bigl(\frac{1}{s}+\frac{2}{t}+2\,\frac{s}{t^2}\Bigr) \nonumber\\
 &+& z\Bigl( 4+4\frac{s}{t}+2\frac{t}{s}\Bigr)
+\frac{1}{2}\frac{t^2}{s}+6\frac{s^2}{t}+5s+4t\Bigr]
\ln\Bigl(-\frac{t}{s}\Bigr)
+ s\Bigl(-\frac{z}{t}+\frac{3}{2}\Bigr)
\nonumber\\
&\times&\ln\Bigl(1+\frac{t}{s}\Bigr)
+\Bigl[\frac{1}{2}\,\frac{z^2}{s}+2\,z\,\Bigl(1+\frac{s}{t}\Bigr)-\frac{11}{4}\,s-2\,t \Bigr]\,
\ln^2\Bigl(-\frac{t}{s}\Bigr)
\nonumber\\
&-&\Bigl[\frac{1}{2}\,\frac{z^2}{t}
-z\,\Bigl(1+\frac{s}{t}\Bigr)+\frac{t^2}{s}
+2\,\frac{s^2}{t}+\frac{9}{2}\,s+\frac{15}{4}\,t \Bigr]\,\ln^2\Bigl(1+\frac{t}{s}\Bigr)
\nonumber\\
&+& \Bigl[\frac{z^2}{t}
-2\,z\,
\Bigl(1+\frac{s}{t}\Bigr) 
+2\,\frac{s^2}{t}+5\,s+\frac{5}{2}\,t
\Bigr]\,\ln\Bigl(-\frac{t}{s}\Bigr)\,\ln\Bigl(1+\frac{t}{s}\Bigr)
\nonumber\\
&-&4\,\Bigl[\frac{t^2}{s}+2\,\frac{s^2}{t}+3\,\Bigl(s+t\Bigr) \Bigr]\,
\Bigl[1
+\textrm{Li}_2\Bigl(-\frac{t}{s}\Bigr)\Bigr]
\nonumber\\
&-&\Bigl[ 2\,\frac{z^2}{t}-4\,z\,\Bigl(1+\frac{s}{t}\Bigr)
-4\,\frac{t^2}{s}
-2\,\frac{s^2}{t}+s-\frac{11}{2}\,t\Bigr]\,\zeta_2\nonumber\\
&-&
 \Bigl[ \frac{t^2}{s}+2\,\frac{s^2}{t}+3\,\Bigl(s+t\Bigr)\Bigr]\,\ln\Bigl(\frac{z}{s}\Bigr)\,
 \ln\Bigl(1+\frac{t}{s}\Bigr)
+\Bigl[ z^2\,\Bigl( \frac{1}{s}
+2\,\frac{s}{t^2}
+ \frac{2}{t}\Bigr)\nonumber\\
&-&z\,\Bigl( \frac{t}{s}+2\frac{s}{t}+2 \Bigr)\Bigr]
\ln\Bigl(\frac{z}{s}\Bigr)
-\Bigl[ z^2\,\Bigl(\frac{1}{s}+\frac{1}{t}\Bigr) +2\,z\,\Bigl(1+\frac{s}{t}\Bigr)
+s
+2\,\frac{s^2}{t}\Bigr]\,\nonumber\\
&\times&\ln\Bigl(\frac{z}{s}\Bigr)\,
\ln\Bigl(1+\frac{z}{s}\Bigr)
+ \Bigl[\frac{z^2}{s}
+4\,z\,\Bigl(1+\frac{s}{t}\Bigr)-\frac{t^2}{s}-4\,\Bigl(s+t\Bigr)\Bigr]\,\nonumber\\
&\times&\ln\Bigl(\frac{z}{s}\Bigr)\,\ln\Bigl(1-\frac{z}{t}\Bigr)
- \Bigl[ z^2\,\Bigl(\frac{1}{s}+2\frac{s}{t^2}+ \frac{2}{t} \Bigr)
-2\, z\,\Bigl(\frac{t}{s}+2\,\frac{s}{t}
+2\Bigr)+\frac{t^2}{s} \nonumber\\
&+&2( s+t)\Bigr]\,
\ln\Bigl(1-\frac{z}{t}\Bigr)
+\Bigl[ \frac{z^2}{t}-2z\Bigl(1+\frac{s}{t}\Bigr)+2\frac{t^2}{s}
+8s
+4\frac{s^2}{t}+7t \Bigr] \nonumber\\
&\times& \ln\Bigl(1-\frac{z}{t}\Bigr)\,\ln\Bigl(1+\frac{t}{s}\Bigr)
-\Bigl[
z^2\,\Bigl(\frac{1}{s}+\frac{1}{t}\Bigr) +2\,z\,\Bigl(1+\frac{s}{t}\Bigr)
+s + 2\,\frac{s^2}{t}
\Bigr]\, \nonumber\\
&\times&\textrm{Li}_2\,\Bigl(-\frac{z}{s}\Bigr)
+ \Bigl[\frac{z^2}{s}+4\,z\,\Bigl(1+\frac{s}{t}\Bigr)
-\frac{t^2}{s}-4\,\Bigl(s+t\Bigr)\Bigr]\,
\textrm{Li}_2\,\Bigl(\frac{z}{t}\Bigr)\nonumber\\
&-& \Bigl[\,\frac{z^2}{t}
-2\,z\,\Bigl(1+\frac{s}{t}\Bigr)
+\frac{t^2}{s}
+ 5\,s+2\,\frac{s^2}{t}
+4\,t\, \Bigr]\,\textrm{Li}_2\,\Bigl(1+\frac{z}{u}\Bigr)
\Bigr\}\nonumber\\
&+&4\,\Bigl(\frac{1}{3}\,\frac{t^2}{s}+\frac{2}{3}\,\frac{s^2}{t}+s+t\Bigr)\,
\ln\Bigl(\frac{2\,\omega}{\sqrt{s}}\Bigr)\,
\Bigl[
\ln\Bigl(\frac{s}{m_e^2}\Bigr)
+\ln\Bigl(-\frac{t}{s}\Bigr)\nonumber\\
&-&\ln\Bigl(1+\frac{t}{s}\Bigr)-1\Bigr]
.
\end{eqnarray}
Table \ref{nums1} and Table \ref{nums2} contain numerical results for small- and large-angle scattering at a variety of 
energy scales. 
We report the QED tree-level prediction, as well as the process-dependent contributions at NNLO
of Eq.~(\ref{TotRemainder}); in other words, we exclude from the tables pure self-energy corrections,
which can be described introducing a running fine-structure constant and were deeply investigated in
the past (see \cite{Arbuzov:2004wp}), and irreducible vertex contributions (see \cite{Kniehl:1988id} and \cite{Burgers:1985qg}).
A complete phenomenological analysis requires also to add the corresponding terms arising from unresolved real 
fermion pair production.

The Standard Model cross sections shown here rely on Born formulae with $Z$ boson and photon exchange a la {\ttfamily Zfitter} \cite{Bardin:1999yd,Arbuzov:2005ma}.
Moreover, although the contributions from electron loops have  been obtained by exact (in $m_e$) evaluation of the Feynman diagrams,
we report here for consistency the approximated results for $m_e<<s, t$. 

We further compare the analytical results of \cite{Actis:2007gi} with those obtained with the dispersion approach and it 
is nicely seen that the former approach the latter in regions where the former are expected to become good approximations.
In both tables, for each fermion flavour, we show the result obtained through
the dispersion-based approach (first line) and the one
coming from the analytical expansion (second line), neglecting ${\cal O}(m_f^2 \slash x)$,
where $x=s,|t|,|u|$. When $m_f^2>x$, the entry is suppressed.
We switch off in the tables the effect of the logarithm containing the energy of soft photons
setting $\omega= \sqrt{s}\slash 2$.

The heavier fermions have less influence on the net result, and the top quark 
decouples nearly completely.
Between 1 and 500 GeV the sum of boxes with factorizable diagrams with muon loops contributes, roughly speaking, at the order of permille to the net pure QED cross section, and the tau lepton contributes less.
The $Z$ resonance distortes this figure, by making the influence of two-loop contributions less influential for large-angle scattering 
where the resonance dominates. 

\begin{table}[ht]\centering
\setlength{\arraycolsep}{\tabcolsep}
\renewcommand\arraystretch{1.1}
\begin{tabular}{|r|r|r|r|r|}
\hline 
$\sqrt{s}$ [GeV] & 1& 10& $M_Z$ & 500\\
\hline 
\hline
{\rm QED~Born} \qquad \quad & 440994  & 4409.94   & 53.0348   & 
 1.76398  \\ 
\hline 
\hline 
{\rm rest}\qquad $e$ &193   & 5.73  & 0.1357  &0.00673   \\
\hline 
$\mu$ & $<$ {\bf 1}  & {\bf 0.42}  & {\bf 0.0408} & {\bf 0.00288}  \\ 
 &  $\times $ & 0.08   &0.0407  & 0.00288  \\
\hline 
$\tau$ &  $<$ {\bf 1}   &  $<$ {\bf 10$^{-2}$}  & {\bf 0.0027} & {\bf 0.00088} \\ 
& $\times$ & $\times$  &-0.0096  & 0.00084 \\
\hline 
$t$ & $<$ {\bf 1}  & $<$ {\bf 10$^{-2}$}  &   $<$ {\bf 10$^{-4}$}  & $<$ {\bf 10$^{-5}$}  \\ 
& $\times$ & $\times$ & $\times$ & $\times$  \\
\hline
\end{tabular}
\caption[]{Numerical values for the differential cross section
in nanobarns at a scattering angle $\theta=3^\circ$, in units of $10^2$;
$M_Z = 91.1876$ GeV. 
Bold-face entries are obtained with dispersion relations.
The energy of the soft photon is chosen to be $\omega= \sqrt{s}\slash 2$.}
\label{nums1}
\end{table}

\begin{table}[ht]\centering
\setlength{\arraycolsep}{\tabcolsep}
\renewcommand\arraystretch{1.2}
\begin{tabular}{|r|r|r|r|r|}
\hline 
$\sqrt{s}$ [GeV] & 1& 10& $M_Z$& 500\\
\hline 
\hline
{\rm QED Born} \qquad \quad & 466537 & 4665.37  &56.1067  &1.86615  \\ 
\hline
\hline
{\rm full~Born} \qquad \quad & 466558 & 4686.27& 1273.2680& 0.85496\\
\hline
\hline
{\rm rest} \quad $e$ & 807  & 14.53   & 0.2706 & 0.01193  \\
\hline 
$\mu$ & {\bf 160}    & {\bf 6.08}  & {\bf 0.1470}  &  {\bf 0.00726} \\ 
      & 153          & 6.08        & 0.1470        & 0.00726   \\
\hline 
$\tau$ & {\bf 2}   & {\bf 1.33} & {\bf 0.0752} 
& {\bf 0.00457}  \\ 
& $\times$ & 1.07   & 0.0752  & 0.00457  \\
\hline 
$t$ & $<$ {\bf 1}  & $<$ {\bf 10$^{-2}$}  & {\bf 0.0005}  & {\bf 0.00043}   \\ 
& $\times$ & $\times$ & $\times$ & -0.00013 \\ 
\hline 
\end{tabular}
\caption[]{Numerical values for the differential cross section 
in nanobarns at a scattering angle $\theta=90^\circ$, in units of $10^{-4}$.
See Table \ref{nums1} for further details.}
\label{nums2}
\end{table}

\section{Summary}
We have evaluated the $n_f=2$ virtual two-loop corrections to Bhabha scattering due to 
fermions with arbitrary mass 
$m_f$ in the limit of vanishing electron mass $m_e$.
We have not combined these (infrared finite) virtual contributions with those arising from irreducible 
vertices; the latter are logarithmically enhanced by terms of order up to $\ln^3(s/m_f^2)$, but are independent of 
$\ln(s/m_e^2)$, being collinear finite.
They have to be assembled with the unresolved real heavy fermion emission, which is known to cancel
the $\ln^3(s/m_f^2)$ and might lead to a suppression of the net effect. 
For the phenomenological analysis we will have also to take into account the effect of
the running of the fine-structure constant.
Concerning the results shown in the tables, the numerical 
contributions do not exceed the per mille level, and depend strongly on the kinematics.
The formulae presented here apply to the leptons $\mu$ and $\tau$, but also to the top quark.
The latter decouples at small energies, but has to be taken into account at the ILC.

The determination of hadronic corrections is in preparation. 

Note added:
\\
After completion of this article, a draft \cite{bon} appeared,
where the authors also study the fermionic corrections to Bhabha
scattering with arbitrary masses of the internal fermions.



\providecommand{\href}[2]{#2}

\end{document}